\documentclass[%
superscriptaddress,
nofootinbib,
twocolumn,
 amsmath,amssymb,
 aps,
pra,
]{revtex4-2}

\usepackage{graphicx}% Include figure files
\usepackage{dcolumn}% Align table columns on decimal point
\usepackage{bm}% bold math
\usepackage{amsmath,amssymb,amsthm,bm,mathtools,amsfonts,mathrsfs,bbm,dsfont} %Useful math packages
\usepackage{graphicx}
\usepackage{braket}
\usepackage{color}
\usepackage{float}
\usepackage[skins,theorems]{tcolorbox}
\usepackage{soul}
\usepackage[paperwidth=210mm,paperheight=297mm,centering,hmargin=1.5cm,vmargin=3.5cm]{geometry}

\usepackage[unicode=true, breaklinks=false, pdfborder={0 0 1}, backref=false, colorlinks=true, linkcolor=blue, citecolor=blue, urlcolor=blue]{hyperref}

\renewcommand{\vec}[1]{\boldsymbol{#1}}

\newcommand{\vR}{\vec{R}}
\renewcommand{\vr}{\vec{r}}
\newcommand{\vk}{\vec{k}}
\newcommand{\vp}{\vec{p}}
\newcommand{\vq}{\vec{q}}
\newcommand{\vg}{\vec{g}}

\newcommand{\ve}{\mathbf{e}}

\newcommand{\la}{\langle}
\newcommand{\ra}{\rangle}

\newcommand{\cL}{\mathcal{L}}

\newcommand{\cT}{\mathcal{T}}

\newcommand{\da}{\dagger}
\newcommand{\eps}{\varepsilon}

\newcommand{\Op}[1]{\hat{#1}}

\newcommand{\oK}{\Op{K}}

\newcommand{\oS}{\Op{S}}
\newcommand{\oT}{\Op{T}}

\newcommand{\oP}{\Op{P}}
\newcommand{\oX}{\Op{X}}

\newcommand{\ox}{\Op{x}}
\newcommand{\oy}{\Op{y}}

\newcommand{\ovr}{\Op{\mathbf{r}}}
\newcommand{\ovp}{\Op{\mathbf{p}}}
\newcommand{\ovR}{\Op{\mathbf{R}}}

\newcommand{\id}{\ensuremath{\mathbbm 1}}

\newcommand{\diff}{\mathrm{d}}

\DeclareMathOperator{\tr}{tr}

\begin{document}

\preprint{APS/123-QED}

\title{Electron-Enabled Nanoparticle Diffraction}

\author{Stefan Nimmrichter}
\email{stefan.nimmrichter@uni-siegen.de}
\affiliation{Naturwissenschaftlich-Technische Fakultät, Universität Siegen, Walter-Flex-Str.~3, 57068 Siegen, Germany}

\author{Dennis R\"atzel}
\affiliation{ZARM, Unversität Bremen, Am Fallturm 2, 28359 Bremen, Germany}
\affiliation{Vienna Center for Quantum Science and Technology, Atominstitut, TU Wien, Stadionallee 2, 1020 Vienna, Austria}

\author{Isobel C. Bicket}
\affiliation{Vienna Center for Quantum Science and Technology, Atominstitut, TU Wien, Stadionallee 2, 1020 Vienna, Austria}
\affiliation{University Service Centre for Transmission Electron Microscopy, TU Wien, Wiedner Hauptstra\ss e 8-10/E057-02, 1040 Wien, Austria}

\author{Michael S. Seifner}
\affiliation{Vienna Center for Quantum Science and Technology, Atominstitut, TU Wien, Stadionallee 2, 1020 Vienna, Austria}
\affiliation{University Service Centre for Transmission Electron Microscopy, TU Wien, Wiedner Hauptstra\ss e 8-10/E057-02, 1040 Wien, Austria}

\author{Philipp Haslinger}
\email{philipp.haslinger@tuwien.ac.at}
\affiliation{Vienna Center for Quantum Science and Technology, Atominstitut, TU Wien, Stadionallee 2, 1020 Vienna, Austria}
\affiliation{University Service Centre for Transmission Electron Microscopy, TU Wien, Wiedner Hauptstra\ss e 8-10/E057-02, 1040 Wien, Austria}

\date{\today}

\begin{abstract}
We propose a scheme for generating high-mass quantum superposition states of an optically pre-cooled, levitated nanoparticle through electron diffraction at its sub-nanometer crystal lattice. 
When a single electron undergoes Bragg diffraction at a free-falling nanoparticle, momentum conservation implies that the superposition of Bragg momenta is imprinted onto the relative coordinate between electron and nanoparticle, which entangles their wavefunctions. By imaging the electron interferogram, one maps the nanoparticle state onto a superposition of Bragg momenta, as if it was diffracted by its own lattice. This results in a coherent momentum splitting approximately 1000 times greater than what is achievable with two-photon recoils in conventional standing-wave gratings. Self-interference of the nanoparticle can thus be observed within drastically shorter free-fall times in a time-domain Talbot interferometer configuration, significantly relaxing source requirements and alleviating decoherence from environmental factors such as residual gas and thermal radiation. Shorter interference times also allow for a recapture of the nanoparticle within its initial trapping volume, facilitating its reuse in many rapid experimental duty cycles. This opens new possibilities for experimental tests of macroscopic quantum effects within a transmission electron microscope.
\end{abstract}

\maketitle

\emph{Introduction.---}Matter-wave interferometry with massive objects challenges our understanding of the quantum-classical transition and is expected to provide necessary experimental insights into quantum gravity \cite{dewitt1957role,anastopoulos2015probing,bose2017spin,Marletto2017,belenchia2018quantum,carlesso2019testing,anastopoulos2020quantum,Danielson2022gravitationally,bose2024massive}. State-of-the-art near-field setups have demonstrated interference for particles up to 
$10^5\,$amu \cite{arndt2014testing, fein2019quantum, pedalino2025probing} 
%$10^4\,$amu \cite{arndt2014testing, fein2019quantum} 
and up to $10^3\,$amu in the time domain \cite{haslinger2013universal}.  While standard quantum theory imposes no fundamental limits on particle size, alternative theories suggest a suppression of macroscopic superpositions, calling for 
%experimental tests 
experiments that could rule out these theories~
\cite{nimmrichter2011testing, bassi2013models,bassi2023collapse}. 
The use of optically cooled, highly controllable nanoparticles presents a promising platform for such experiments \cite{delic2020groundState, perdriat2021_NVcenter_charged_review, piotrowski2023_2D_groundstate_cooling, perdriat2023spin_chargedNS,bykov2024nanoparticle_charged,jin2024_NVcenter_charged,rossi2024quantum,muffato2024coherent}. Although numerous proposals have highlighted strategies for high-mass interferometry \cite{romeroisart2011large,nimmrichter2011concept,bateman2014near,wan2016Free,romero-isart2017coherent,bose2017spin,kaltenbaek2023maqro,roda2024macroscopic,wardak2024nanoparticle,steiner2024pentacene}, experimental realization still remains a challenge and hinges on innovative techniques to realize coherent matter-wave diffraction, reduce environmental disturbances, lower the  interference time, and improve experimental stability and repetition rates. 
While most approaches rely on optical fields as diffractive elements for matter waves, we propose an Electron-Enabled Nanoparticle Diffraction (END) scheme that exploits the underlying atomic structure of a nanoparticle through high-resolution electron imaging. In effect, the particle can be made to diffract off its own crystal lattice, resulting in drastically lower interference times and experimental requirements than other approaches. Table \ref{tab:mass-talbot_time} lists the required free-fall times and distances for nanoparticles of various sizes to interfere, given the exemplary diffraction period of $192\,$pm.

\begin{table}
    \centering
    \scriptsize
    \renewcommand{\arraystretch}{1.5}
    \begin{tabular}{|p{2cm}|>{\centering}p{2cm}|>{\centering}p{2cm}|>{\centering\arraybackslash}p{2cm}|}
        \hline
        \multicolumn{4}{|c|}{\normalsize \textbf{High-Mass Interferometry d=192 pm}} \\ \hline
         {\textbf{System}} & \textbf{Mass} &  {\textbf{Talbot Time}} & {\textbf{Free Fall}}  \\ \hline
          Bateman et al. 2014 \cite{bateman2014near} & $\mathbf{1 \cdot 10^6}$ \textbf{amu} & $9.2 \cdot 10^{-8}$ s & $1.7 \cdot 10^{-13}$ m \\ \hline
         Deli\'c et al. 2020 \cite{delic2020groundState} ($\mathrm{SiO_2}$) & $\mathbf{2 \cdot 10^9}$ \textbf{amu} & $1.8 \cdot 10^{-4}$ s & $6.7 \cdot 10^{-7}$ m \\ \hline
         Bykov et al. 2024 \cite{bykov2024nanoparticle_charged} ($\mathrm{SiO_2}$) & $\mathbf{2 \cdot 10^{10}}$ \textbf{amu} & $1.8 \cdot 10^{-3}$ s & $6.7 \cdot 10^{-5}$ m \\ \hline
         Jin et al. 2024 \cite{jin2024_NVcenter_charged} (diamond) & $\mathbf{1 \cdot 10^{11}}$ \textbf{amu} & $9.2 \cdot 10^{-3}$ s & $1.7 \cdot 10^{-3}$ m \\ \hline
         Perdriat et al. 2021 \cite{perdriat2021_NVcenter_charged_review} (diamond) & $\mathbf{7 \cdot 10^{11}}$ \textbf{amu} & $6.5 \cdot 10^{-2}$ s & $8.2 \cdot 10^{-2}$ m \\ \hline
    \end{tabular}
    \caption{Talbot times and typical free-fall distances for the interference of different nanoparticle masses considered in recent publications \cite{bateman2014near,piotrowski2023_2D_groundstate_cooling,delic2020groundState,perdriat2023spin_chargedNS,bykov2024nanoparticle_charged,jin2024_NVcenter_charged,perdriat2021_NVcenter_charged_review} at the exemplary grating period $d=192\,$pm considered here. We require that the particle free-falls over twice the Talbot time to observe interference fringes. The short free-fall distances facilitate a recapture of the particle within a single trap volume.}
    \label{tab:mass-talbot_time}
\end{table}

The END approach takes inspiration from the first matter-wave experiments by Davisson and Germer \cite{Davision_Germer_e-diff1927} demonstrating elastic Bragg diffraction of electrons at the atomic lattice structure of a crystalline specimen. If the Bragg condition is met for a given period $d$ of lattice planes, the electron wavefunction splits into a superposition of diffraction components separated by multiples of the (mass- and velocity-independent) Bragg momentum $p_d = h/d$, where $h = 2\pi\hbar$ denotes Planck's constant. These Bragg components would show up as distinct bright spots on a detection screen in the far field (Fourier plane) or they could be filtered and refocused to form $d$-periodic interference fringes on a farther image plane.
Crucially, momentum conservation dictates that for each Bragg momentum imparted on the electron, the crystal particle receives a recoil of equal magnitude and opposite sign. In principle, Bragg diffraction in a selected spatial direction can thus generate EPR-like entanglement as it maps product wavefunctions of electron and particle, $\psi(x) \Psi(X)$, to a superposition of states that are momentum-shifted in the \textit{relative} coordinate $x-X$, 
\begin{equation}
    \psi(x)\Psi(X) \mapsto \sum_n f_n \exp \left[ \frac{i}{\hbar} n p_d (x-X) \right]\psi(x)\Psi(X).
\end{equation}
By detecting the electron position $x$ on the image plane, where the Bragg orders interfere, one effectively erases which-way information and projects the particle state onto a phase-shifted superposition of Bragg-diffracted components,
\begin{equation}\label{eq:gratingTrafo_psi}
    \Psi_x(X) \propto \sum_n f_n \exp \left[ \frac{i}{\hbar} n p_d (x-X) \right]\Psi(X),
\end{equation}
up to normalisation. The outcome looks as if the particle had been diffracted off a grating with a period $d$ given by the particle's own crystal lattice. Interference between the components can be observed after an additional free evolution time on the order of the Talbot time, $T_M = Md^2/h$, which grows with the particle mass $M$. 

The coherent recoils and their said implications are typically of no relevance in Bragg spectroscopy or in electron imaging of massive (often fixed) crystal samples. However, they can have a sizable effect on levitated, motionally cooled nanoparticles, which have become available only recently in a mass range of $M \sim 10^6 - 10^{12}\,$ amu \cite{piotrowski2023_2D_groundstate_cooling,delic2020groundState,perdriat2023spin_chargedNS,bykov2024nanoparticle_charged,jin2024_NVcenter_charged,perdriat2021_NVcenter_charged_review}. 
In fact, the Bragg momenta here are about a thousand times greater than the diffraction momenta of optical grating structures in existing proposals for high-mass interferometers \cite{romeroisart2011large,nimmrichter2011concept,bateman2014near,kaltenbaek2023maqro,wardak2024nanoparticle}, which implies a drastic reduction of the required free evolution times by a million for a given mass.

Having said that, the realisation of the effective grating transformation \eqref{eq:gratingTrafo_psi} requires a precisely timed coherent single-electron pulse to hit the particle and a post-selection scheme that filters out all Bragg components other than the multiples of $p_d$ in the electron state. Moreover, since the superposition in \eqref{eq:gratingTrafo_psi} varies in phase with the measured electron position $x$, the measurement resolution must be better than the Angstrom-sized period $d$. Otherwise, the interference contrast will be reduced. 
Transmission electron microscopy (TEM) is expected to meet these challenges in the near future; current TEMs already enable spatial resolutions $<$50 pm \cite{kisielowski2008HRTEM, jin2017atomic, urban2023progressCTEM, Ishikawa2023} with highly coherent illumination \cite{tsujino2022coherence}, imaging on the micrometer scale \cite{Reimer2008}, temporal control on the sub-picosecond scale \cite{Lobastov2005,Zewail2006,Vanacore2018_atto}, few-$\mu$rad or better resolution of diffraction angles \cite{mori2021SmAED}, and near-unitary detection efficiency \cite{llopart2022timepix4}.

\begin{figure*}
    \includegraphics[width=16cm]{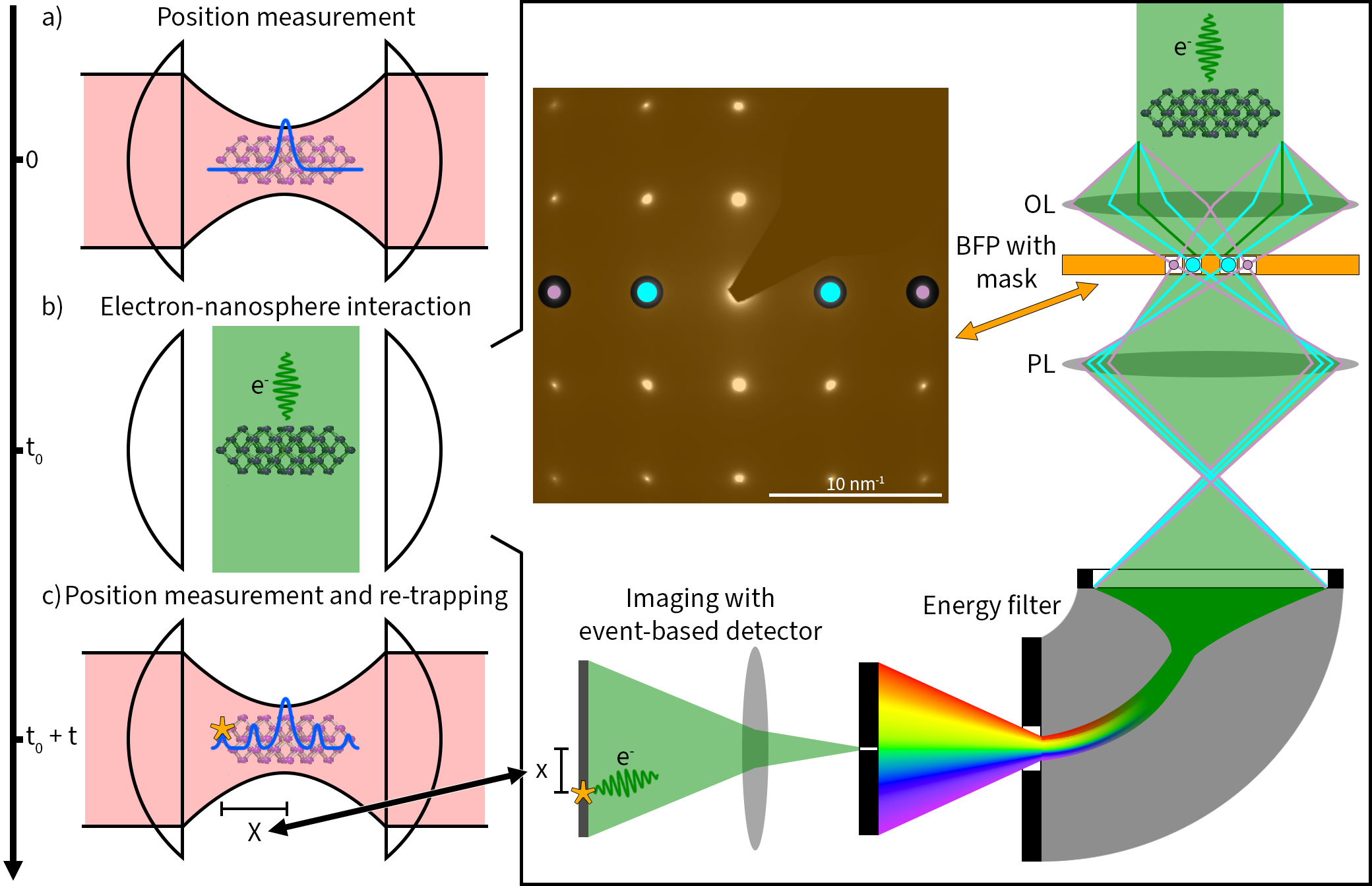}
    \caption{Proposed scheme of electron-enabled nanoparticle diffraction consisting of three steps: (a) a nanoparticle of mass $M$ is cooled close to its motional ground state in, e.g., an optical trap, acting as a highly localised matter wave source released into free fall when the trap is switched off. After a sufficient buildup of coherent delocalisation over the time $t_0$, (b) a triggered single-electron wave packet impinges and Bragg-diffracts off the particle's lattice structure. As each imparted Bragg momentum comes with an equivalent recoil, the wavefunctions of particle and electron are now entangled. The electron passes a mask on the back focal plane (BFP) of an objective lens (OL), which selects Bragg orders corresponding to a lattice period $d$ (Inset: experimental diffraction image of a silicon crystal, illustrating the intensity distribution when aligned along the relevant zone axis, and the selection of Bragg orders of periodicity $d=192\,$pm). The selected orders are recombined on an image plane by the projector lens (PL) system and an energy filter is used to exclude inelastic events. The detector at the image plane records the electron's position $x$. This effectively maps the particle wavefunction into an $x$-dependent superposition of Bragg momenta. They are allowed to interfere over another free-fall time $t$, before (c) the position $X$ is measured and the particle re-trapped. Given free-fall times of at least one Talbot time, $t,t_0 \gtrsim T_M$, stable interference fringes of magnified period $D=d(1+t/t_0)$ in the relative coordinate $X-xD/d$ will form over many valid repetitions (i.e., whenever the electron is detected).} 
    \label{fig:expScheme}
\end{figure*}

\emph{Proposed experiment.---}% 
The proposed experimental sequence illustrated in Fig.~\ref{fig:expScheme} adopts the basic interference scheme of Ref.~\cite{bateman2014near}, but with significant modifications. The experiment is conducted within an ultra-fast transmission electron microscope and utilizes pulsed beams of single electrons that can be imaged with sub-atomic resolution. 
In the first stage (a), a nanoparticle is captured and cooled, e.g., in an optical dipole trap \cite{lindner2024hollow_NS_loading, delic2020groundState} or ion trap \cite{bykov2024nanoparticle_charged, dania2022position}. The position of the particle is measured and, e.g., feedback or cavity cooling is applied to further localise the nanoparticle state. The trap thus serves as a point-like matter-wave source. 

After release from the trap, the nanoparticle evolves freely for a time $t_0$, before (b) a triggered single-electron pulse hits the particle and diffracts off its crystal lattice. The momentum state of the electron thus splits into distinct diffraction components separated by discrete Bragg momenta, each of which imparts an equal momentum recoil on the particle in the opposite direction. A chosen Bragg momentum $p_d$ and multiples thereof are then post-selected with the help of a pinhole mask placed in the back focal plane (BFP) of an objective lens (OL), which blocks the undiffracted electron state and any Bragg component other than $n p_d$ in a given spatial direction ($x$-axis) \cite{danz2021ultrafast}. To this end, one must align the particle's crystal lattice accordingly. 
The selected Bragg components are allowed to interfere on an image plane further down the beam line, where the electron position $x$ is measured. At perfect resolution, this approximately projects a pure centre-of-mass state onto the post-measurement state \eqref{eq:gratingTrafo_psi}, equivalent to the effect of a diffraction grating of period $d$. 
See Appendix \ref{app:gratingTrafo} for a detailed derivation. 
The nanoparticle can be assumed at rest during this process, which takes a few nanoseconds.
Finally, the nanoparticle evolves for another time $t$ before (c) its centre-of-mass position $X$ is measured and it is recaptured in the original trap for the next run of the sequence. We discard runs in which the electron is not detected. Over many cycles, the measured position distribution exhibits interference fringes shifted with respect to the electron position $x$, which can be calculated using standard phase-space methods; see Appendix \ref{app:phasespace_calc}.

\begin{figure*}
    \centering
    \includegraphics[width=1.0\linewidth]{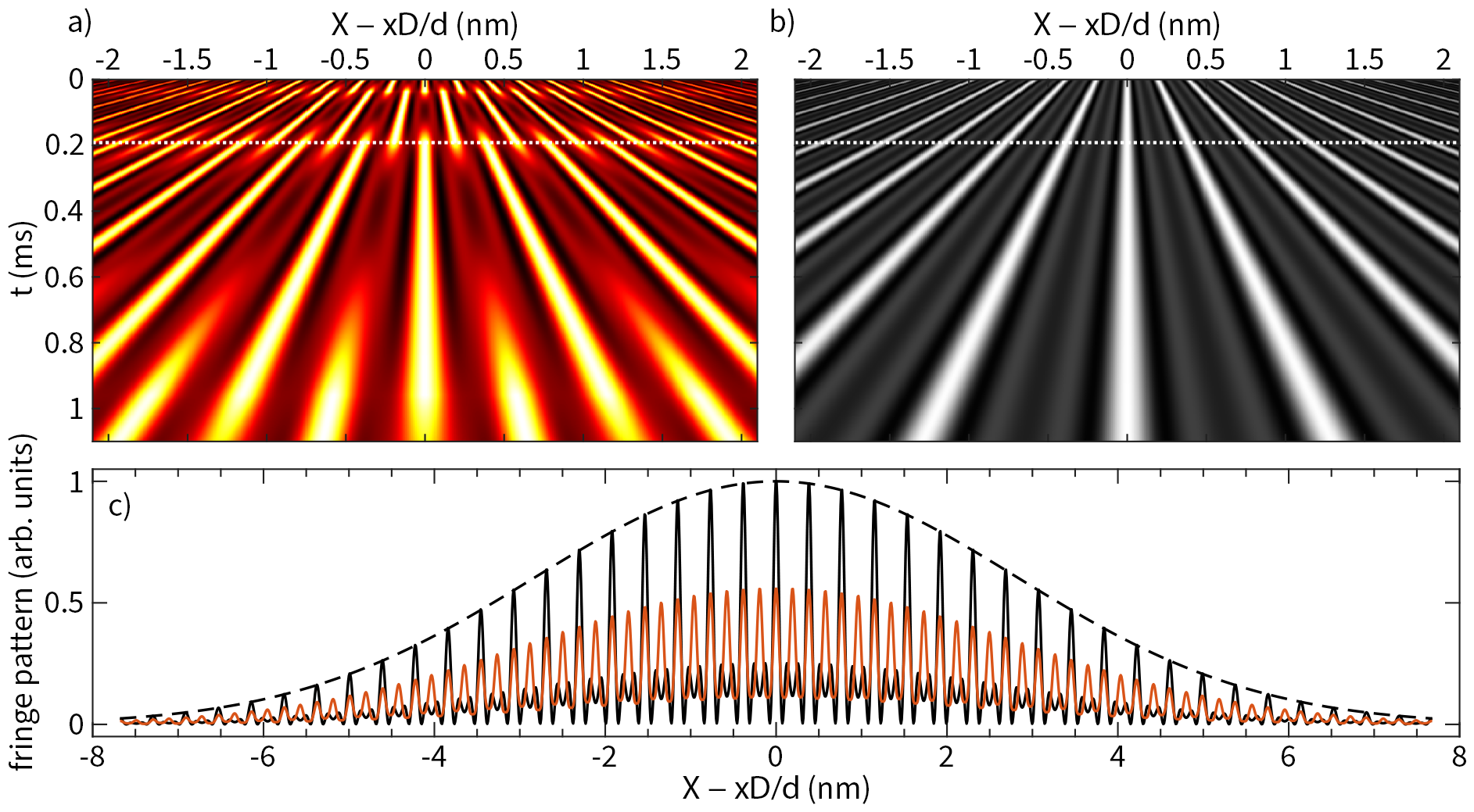}
    \caption{Exemplary fringe patterns for an aligned silicon nanocrystal of mass $M=2\times 10^9\,$amu, based on electron Bragg diffraction at $\pm(1\bar{1}0), \pm(2\bar{2}0)$ and plotted against the relative coordinate $X-xD/d$ with respect to the detected electron position $x$. The particle is released and expands freely for the time $t_0 = T_M = 192\,\mu$s before the electron diffraction. In (a), we plot the position distribution of the particle as a function of time $t$ after diffraction, normalised to its maximum value at each $t$. In (b), we show the hypothetical shadow pattern for a classical particle transmitted by a classical aperture. (c) Position distributions at $t=T_M$ [dotted lines in (a) and (b)] corresponding to quantum interference (red solid) and classical shadow fringes (black solid), in units relative to the maximum of the classical pattern. The dashed line indicates the Gaussian distribution of a freely evolved particle without diffraction.}
    \label{fig:carpet}
\end{figure*}

Figure \ref{fig:carpet} depicts exemplary interference fringe patterns we predict assuming perfect measurement resolution. The pattern is given relative to the measured electron position $x$ and would be smeared out accordingly at finite resolution. In our case study, we consider a silicon nanoparticle of mass $M=2\times 10^9\,$amu and the trap parameters of Ref.~\cite{delic2020groundState} ($305\,$kHz trap frequency, $12\,\mu$K center-of-mass temperature), which results in a Gaussian source state of standard deviation $\sigma_X \approx 3.8\,$pm, close to the motional ground state.
Our mask shall consist of 4 pinholes selecting the multiples $n=\pm1,\pm2$ of the Bragg order specified by the Miller indices $(1\overline{1}0)$ in primitive-cell representation [i.e., $(20\overline{2})$ in the conventional cubic-cell notation]. The associated grating period is $d=192\,$pm, which demands that the pinholes be a few micrometres in size \cite{danz2021ultrafast}. To ensure spatial coherence of the nanoparticle over one period, we let it evolve for one Talbot time before the electron interaction, $t_0 =T_M \approx 192\,\mu$s. In (a), we show the interferogram of the recaptured nanoparticle after varying times $t$. The fringe period magnifies geometrically as $D = d(1+t/t_0)$. Notice that the existence of fringes does not certify the quantum wave nature of the nanoparticle. Treating the electron interaction stage as a classical grating aperture would result in the shadow pattern (b), and one cannot claim quantum interference whenever the patterns (a) and (b) coincide. Panel (c) compares the fringe patterns at $t_0,t=T_M$, where they differ the most. To reveal the quantum fringes, the measurement resolution must be better than $d/4 =48\,$pm for the electron and $D/4 = 96\,$pm for the nanoparticle, which are within reach \cite{ Ishikawa2023,millen2020optomechanics, dania2022position, bonvin2024hybrid}. 
The short total interference time $t+t_0 = 384\,\mu$s amounts to a free-fall distance of less than a micrometre, which allows for recapture and thus a fast duty cycle, as we will briefly discuss next.   

Our exemplary analysis is carried out for a typical electron energy of $300\,$keV and a silicon nanoparticle in the shape of an oblate spheroid of radius $R_M=109\,$nm and thickness $2b_M=60\,$nm. This amounts to the mass of a particle for which ground-state cooling was already achieved \cite{delic2020groundState}, but our scheme can also be operated with larger particles. 
The electrons shall have a Gaussian transverse profile with a $115\,$nm half-width-half-max spot size. From this, we estimate the probability to detect the Bragg-filtered electron as $\Pr_{\rm det} \approx 0.1\%$. Temporal triggering suppresses unwanted detection events due to, e.g., cosmic rays. 
The duty cycle of our proposed scheme is mainly constrained by the free evolution time, which can be as short as two Talbot times (see Fig.~\ref{fig:carpet}). Particles lighter than $2.5\cdot 10^{9}\,$amu would fall less than $1\,\mu$m and could be easily re-trapped within the same potential for reuse in subsequent experiments, allowing for experimental repetition times of about $1\,$ms. Consequently, considering $\Pr_{\rm det}$ and the Poissonian nature of the electron production process, we estimate approximately 1000 successful experimental runs within one hour.

\emph{Discussion.---}% 
The proposed electron-enabled nanoparticle diffraction scheme represents a viable route towards high-mass matter-wave experiments with levitated nanoparticles at the mass scale of $10^9\,$amu and beyond.
Leveraging the precise control and high spatio-temporal resolution of present-day transmission electron microscopy, the END scheme exploits the coherent recoil caused by Bragg diffraction off atomic structures, which entangles the wavefunctions of electron and nanoparticle. Upon electron detection, the particle self-interferes within a short free-fall time of less than a millisecond (at $10^9\,$amu); see Table~\ref{tab:mass-talbot_time}.
This not only relaxes the experimental constraints associated with high-vacuum conditions and black-body radiation, but it also facilitates a reliable recapture of the nanoparticles and fast experimental duty cycles. 

Our exemplary case study with $M=2\times 10^9\,$amu and an interference time $t = 1\,$ms could reach a logarithmic macroscopicity as high as $\mu \approx 16.3$---one order of magnitude above the status quo  \cite{pedalino2025probing,schrinski2020quantum}. This is based on an empirical measure that compares the `size' of quantum superposition states achieved in experiments with mechanical degrees of freedom \cite{nimmrichter2013macroscopicity}; see App.~\ref{app:macroscopicity} for details.  
%Going further, 
Adopting a recent idea for entangling electrons through a quantum scatterer \cite{Ruimy2024manybody},
one could envisage whole sequences of consecutive electron interactions for multi-particle entanglement and interference experiments within shallow nanoparticle traps.

\acknowledgments
The authors thank Markus Arndt, Peter Schattschneider and Paul Hamilton for fruitful discussions. 
PH thanks the Austrian Science Fund (FWF): Y1121, P36041, P35953 and the FFG-project AQUTEM. DR acknowledges support by the Deutsche Forschungsgemeinschaft (DFG, German Research Foundation) under Germany’s Excellence Strategy – EXC-2123 QuantumFrontiers – 390837967.

%\bibliography{bib}

%\newpage
\appendix 

\section*{End matter}

For the experimental case study discussed in the main text, we make the following assumptions and present estimates about the most relevant systematic effects.
We assume that the electrons are only scattered once and elastically, neglecting both multi-scattering and inelastic scattering. 
Multi-scattering can either enhance or suppress the amplitudes of certain Bragg peaks, calling for a detailed analysis based on the actual shape and size of the nanoparticle in a realisation of our proposal.
Given an inelastic mean free path in silicon of about $180\,$nm \cite{lee2002SiIMPF}, unwanted decoherence due to inelastic scattering should be negligible in our case study. 
To mitigate decoherence for larger particles, the electrons can also be post-selected according to the amount of energy they have lost. Modern TEM setups with electron energy filters allow for an energy filtered imaging resolution of  less than 250 meV \cite{koch2006EFTEM, ruiz_caridad2022cefid}. 

We further assume that the orientation of the nanoparticle is controlled. This could be achieved, for instance, if the particle exhibits an electric dipole moment and is subjected to a homogeneous electric field;  
nanoscale rotational control schemes have been proposed and tested recently \cite{stickler2021quantum,schrinski2022interferometric,kamba2023nanoscale}. 
The effect of such a static field on the electrons can be compensated using coils that deflect the beam in the opposite direction.
If the orientation of the nanoparticle is not controlled, our scheme still works, albeit with a reduced scattering rate into the chosen Bragg peaks and a loss of contrast due to imperfect Bragg peak selection. 
See Appendix~\ref{app:rotationAverage} for an exemplary assessment of the visibility loss due to imperfect alignment.

Localized charges attached to the nanoparticle (e.g., remaining surface charges or charges created by the electron beam) will induce an additional electrostatic deflection of the electrons by an angle of the order of $e^2/(2\pi\eps_0 m_e\gamma v^2 b)$, where $\gamma=(1-v^2/c^2)^{-1/2}$ is the Lorentz factor, $c$ is the speed of light and $b$ is the impact parameter of the electron with respect to the localized charge. For $b=1\,$nm and a kinetic energy of $300\,$keV, the deflection angle is of the order of $10^{-5}\,$rad and can be safely neglected.
Magnetic dipole interactions with unpaired electron spins (e.g., dangling bonds) are even weaker~\cite{Haslinger_2024}.

Another potential risk for the experiment could be the electron-induced sublimation of silicon atoms, changing the mass and potentially the shape of the nanoparticle. However, due to the relatively low electron dose proposed for the successful implementation of the experiment ($10^6$ electrons per $3.7 \times 10^4\,$nm$^2$ cross-section area), the chance of a sputtering event is negligible \cite{Egerton2010sputtering}. Back-scattering of a $300\,$keV electron would impart at most $0.2\,$mm/s onto the nanoparticle, which does not lead to particle loss.

Optical trapping of nanoparticles can result in high internal temperatures due to photon absorption \cite{hebestreit2018measuringTemp}, which may cause a gradual rearranging of atoms in the crystal and thus deformation. This problem could be mitigated by employing a suitable dielectric material with low absorption or by operating with a charged nanoparticle in an ion trap \cite{jin2024_NVcenter_charged,bykov2024nanoparticle_charged}.

Additionally, incorporating optical or ion traps into the confined space between the polepieces of a TEM is a non-trivial, but feasible, task. 
The coupling of light into or out of the TEM sample region has been accomplished in many different ways for cathodoluminescence or photon-induced near-field electron microscopy, with the use of fibre-optics \cite{feist_cavity-mediated_2022, Scheucher2022} or miniaturized free-space mirrors \cite{yamamoto_characterization_1990, zagonel_visualizing_2012, yanagimoto_time-correlated_2023, preimesberger_experimental_2025, varkentina_cathodoluminescence_2022} allowing the possibility of free-space diffraction-limited optical imaging of the sample region as well as the measurement of temporal correlations. High intensity laser cavity systems have also been successfully integrated \cite{schwartz2019laser}, demonstrating the feasibility of establishing strong optical fields inside the TEM. 
Given the experience with such optical systems as well as the growing interest in developing large-gap polepieces \cite{maekawa_development_2025, taheri_current_2016, sharma_perspective_2024}, we anticipate that it will be possible to incorporate an optical or ion trap into a TEM.

\emph{Decoherence estimates.---}% 
Massive neutral particles are subject to decoherence, mainly caused by emission of thermal radiation and by collisions with residual gas particles \cite{hornberger2004theory}. The result is an exponential reduction of fringe visibility with an exponent $\Gamma_{\rm dec} (t+t_0)$, proportional to the total interference time and an effective decoherence rate that depends on the material properties, the mass $M$, and the fringe period $d$. Compared to other high-mass interference schemes, the interference time and fringe period are much shorter here, alleviating the impact of decoherence. 
For a simple conservative estimate, we extrapolate from the experiment proposed in Ref.~\cite{bateman2014near} for a silicon particle of $10^6\,$amu, with $10^2\,$ms interference time, and $10^3\,$nm fringe period. Here, the mass is more than a factor $10^3$ greater, the interference time $10^3$ smaller, and the fringe period more than $10^3$ smaller. 
For the case of gas collisions, $\Gamma_{\rm dec}$ is conservatively estimated by the collision rate, which scales roughly like $M^{2/5}$. Decoherence is therefore \textit{reduced} by about $10^{-9/5}$ compared to Ref.~\cite{bateman2014near}.
Decoherence by thermal emission can be modeled as a diffusion process due to the weak recoil of blackbody photons at $\mu$m-sized wavelengths. In this case, $\Gamma_{\rm dec} \propto M d^2$, which yields a decoherence effect suppressed by $10^{-6}$ relative to Ref.~\cite{bateman2014near}. Hence, decoherence should be irrelevant for vacuum pressures of 
$10^{-9}\,$mbar, which is achievable in modern TEMs~\cite{Reidy2023UHV}, and internal particle temperatures below their melting point.

Dephasing due to charging of the nanoparticle can also be safely neglected. The dipole trap is placed in between the pole pieces of the TEM, providing a spherical volume with a radius $r=2\,$mm of vacuum to decouple the nanoparticle from the environment. Charging of the nanoparticle with an additional electron will cause an electrostatic deflection due to mirror charges of $\Delta s=(e^2/4\pi \eps_0 r^2)t^2/2M$ which amounts to $\Delta s\approx 1$\,pm within two Talbot times, much smaller than the fringe spacing $d$. 

\emph{Spatial resolution.---}% 
An important technical requirement for the proposed experiment is to resolve the interference fringes of the electron and the nanoparticle. On the electron side, our proposal demands a spatial resolution of about $d/4 =48\,$pm.
The imaging resolution is mainly limited by lens aberrations or system instabilities \cite{kisielowski2008HRTEM}. Chromatic aberrations can be minimised using chromatic aberration correctors \cite{haider_information_2010, borrnert_chromatic_2018}, or by reducing the energy spread present in the image via incident beam monochromation \cite{tiemeijer_using_2012} or post-selection energy-filtering \cite{Egerton_1997}. Recent progress in the correction of spherical aberrations resulted in a 3rd-order aberration coefficient as small as $C_s = 350\,$nm, which theoretically supports 
a resolution down to $d_s = \lambda (C_s/2\lambda)^{1/4} = 34\,$pm \cite{kisielowski2008HRTEM}. 
In practice, the resolution is currently still limited by higher-order aberrations to about $50\,$pm, which will continue to improve in ongoing efforts.

On the nanoparticle side, one must be able to resolve distances $D/4 = 96\,$pm upon recapture. 
State-of-the-art cavity-enhanced schemes allow position measurements at the $10^{-14} {\rm m/\sqrt{Hz}}$ level \cite{millen2020optomechanics}, while more conventional setups have demonstrated a position sensitivity below $100$~pm within less than 20~$\mu$s \cite{tseng2025darksearch}.

\newpage 

\section{Effective grating transformation}\label{app:gratingTrafo}

Consider the Bragg diffraction of a pulsed single-electron beam at a single silicon crystal, determined by a scattering operator $\oS = \id +i\oT $ that maps an incoming wave function $|\psi_{\rm in}\ra$ into $|\psi_{\rm out}\ra = |\psi_{\rm in}\ra + i\oT |\psi_{\rm in}\ra$. We operate in the paraxial regime of high kinetic energy ($E_0=eU$), in which the de Broglie wavelength, $\lambda = h c/\sqrt{E_0(2m_ec^2+E_0)}$, is much smaller than the lattice constant $a$. Then, given beam propagation along the $z$-axis and a not too large crystal volume, the electron will undergo a single small-angle scattering transformation at the crystal's periodic lattice of atoms, as described by the eikonal approximation \cite{Reimer2008},
\begin{equation}\label{eq:S}
    \oS \approx \exp \left[ i \sum_{hk\ell} f_{hk\ell} e^{i\vg_{hk\ell}\cdot \ovr_{\perp}} \overline{\varrho} (\ovr_{\perp}) \right].
\end{equation}
Here, $\ovr_{\perp} = (\ox,\oy)$ denotes the transverse position operator of the electron relative to the crystal's centre of mass. The sum consists of Bragg momentum displacements by reciprocal lattice vectors $\vg_{hk\ell}$ of length $g_{hk\ell} = 2\pi/d_{hk\ell}$, specified by Miller indices $(hk\ell)$. The displacements are weighted by the amplitudes
\begin{equation}
    f_{hk\ell} = F_{hk\ell} \left( 1 + \frac{E_0}{m_e c^2}\right) \frac{2Z_{\rm Si}^{2/3}a_{\rm Si} \lambda}{1+(2\pi a_{\rm Si}/d_{hk\ell})^2} , \label{eq:fhkl}
\end{equation}
assuming single-atom scattering in the Wentzel model at small scattering angles, with atomic parameters $Z_{\rm Si} = 14$, $a_{\rm Si} = a_0/Z_{\rm Si}^{1/3}$, and $a_0$ the Bohr radius. The term $F_{hk\ell}$ is the structure factor of the lattice unit cell. Of course, the Bragg diffraction in \eqref{eq:S} is restricted to beam trajectories that penetrate the crystal volume $V\gg a^3$, which is taken into account by the homogeneous, trajectory-averaged density of unit cells $\overline{\varrho}(\vr_{\perp}) = \int\diff z\, \varrho(\vr_{\perp}+z\ve_z)$. 
For an oblate ellipsoid of thickness $2b_M$ and radius $R_M$ containing $N_{\rm cell}$ unit cells, we get
\begin{equation}\label{eq:rhoeff_sphere}
    \overline{\varrho}(\vr_{\perp}) = \frac{N_{\rm cell}}{V}\int \diff z \, \Theta \left(1 - \frac{r_{\perp}^2}{R_M^2}-\frac{z^2}{b_M^2} \right) 
\end{equation}
with $\Theta$ the Heaviside function and $V=4\pi b_MR_M^2/3$.

Silicon has a diamond cubic crystal structure with lattice constant $a=543\,$pm, a primitive unit cell of $2$ atoms, and a structure factor $F_{hk\ell} = 2\cos[(h+k+\ell)\pi/4]$; any Bragg order for which $(h+k+\ell)/2$ is an odd integer is therefore kinematically forbidden.

As the electron propagates freely to the far-field, or back focal plane of an objective lens, its wavefunction separates into spatially distinct Bragg components. A transmission mask of circular pinholes allows us to select a subset of these components and block all the others. We can describe this by multiplying an aperture function $M(\vp_{\perp})$ to the state in momentum representation, $|\psi_{\rm out} \ra \to |\psi_{\rm sel} \ra = M(\ovp_{\perp}) |\psi_{\rm out}\ra$. Here we consider a linear configuration of pinholes separated along the space-fixed $x$-axis by multiples of a fixed diffraction momentum $2\pi\hbar/d$, 
\begin{equation}\label{eq:M_mask}
    M(\vp_{\perp}) = \sum_{n\neq 0} M_0 \left( \left|\vp_{\perp}- n \frac{2\pi\hbar}{d}\ve_x\right|\right),
\end{equation}
where we choose a $d=d_{h_0k_0\ell_0}$ with a sizeable scattering amplitude $f_{h_0k_0\ell_0}$. Each pinhole thus selects a diffraction component of the scattered electron state with matching Bragg momentum, $g_{hk\ell} = 2\pi n/d$, provided there is one pointing in $x$-direction for the given crystal orientation. Since we assume that the undiffracted component ($n=0$) is blocked, $M(\ovp_{\perp})|\psi_{\rm in}\ra = 0$, the overall transformation can be written as $|\psi_{\rm sel} \ra = i M(\ovp_{\perp})\oT |\psi_{\rm in}\ra$. Expanding the scattering operator \eqref{eq:S} to linear order, we can further simplify $\oT \approx \sum_{hk\ell} f_{hk\ell} e^{i\vg_{hk\ell}\cdot \ovr_{\perp}} \overline{\varrho}(\ovr_{\perp})$. 
For our case study, we pick $(h_0k_0\ell_0)$ from the $\{1\bar{1}0\}$-family of lattice planes with a fairly large period, $d = a/2\sqrt{2} = 192\,$pm, and we transmit the Bragg peaks up to the 
second order, $n=\pm 1,\pm 2$. 
%third order, $n=\pm 1,\pm 2,\pm 3$. 
Expressed in terms of the conventional cubic unit cell, the chosen family corresponds to $\{H_0K_0L_0\} = \{20\bar{2}\}$ \cite{Jackson1991Handbook}.

The electron then propagates further (either freely or by means of lens imaging) to the image plane where the selected Bragg orders interfere and form a fringe pattern of period $d$. Measuring the electron in a position-resolving detector amounts to projecting the wavefunction $|\psi_{\rm sel}\ra$ onto a position eigenstate $|\vr_{\perp}\ra$ corresponding to the registered outcome. This assumes a spatial resolution much better than $d$. 

So far, we have treated the crystal as a fixed motionless object. In our setting however, both the centre-of-mass position $\vR$ and the orientation $\Omega$ of the crystal are dynamical (quantum) variables that enter the scattering transformation $\oT$. We can re-introduce them explicitly by noting that the electron position in $\oT$ is measured relative to $\vR_{\perp} = (X,Y)$ and the body-fixed reciprocal lattice vectors $\vg_{hk\ell}$ are rotated by $\Omega$ with respect to the space-fixed frame,
\begin{equation}\label{eq:T1}
\oT(\vR_{\perp},\Omega) = \sum_{hk\ell} f_{hk\ell} e^{i R(\Omega)\vg_{hk\ell} \cdot (\ovr_{\perp}-\vR_{\perp})} \overline{\varrho}(\ovr_{\perp}-\vR_{\perp}).  
\end{equation}
Here, $R(\Omega)$ denotes a rotation matrix, which one can parametrise in terms of Euler angles $(\alpha\beta\gamma)$, for example.

Now let $|\Psi_{\rm cm}\ra$ be the centre-of-mass wavefunction of the nanoparticle upon the scattering event and let $|\Omega\ra$ be an arbitrary orientation state of the lattice. The scattering operator $\oT$ acts on the product state of electron and crystal. Conditioned on the detected electron position $\vr_{\perp}$, the Bragg scattering event transforms the motional state of the crystal as $|\Psi_{\rm cm}\ra|\Omega\ra \to \oK(\vr_{\perp},\Omega) |\Psi_{\rm cm}\ra|\Omega\ra$, where
\begin{equation}\label{eq:K1}
    \oK(\vr_{\perp},\Omega) = \la \vr_{\perp}|iM(\ovp_{\perp}) \oT(\ovR_{\perp},\Omega)|\psi_{\rm in}\ra .
\end{equation}
Here we neglect the electron's time of flight to the detector, which is much shorter than any motional time scale of the crystal. 
The transformation of the reduced centre-of-mass state is then obtained by averaging over a distribution $\mu (\Omega)$ of orientations the crystal assumes at the moment of scattering, given an uncorrelated incoherent mixture of orientations. For a general (pure or mixed) state $\rho_{\rm cm}$, the transformation reads as
\begin{equation}\label{eq:map_cm1}
    \rho_{\rm cm} \mapsto \int \diff^3 \Omega \, \mu(\Omega) \oK(\vr_{\perp},\Omega) \rho_{\rm cm} \oK^\da (\vr_{\perp},\Omega),
\end{equation}
which can be seen as the Kraus representation of a completely positive (but not trace-preserving) map on the centre-of-mass state.
We will see that it acts as a partially coherent grating transformation that can lead to de Broglie self-interference of the crystal, depending on the spread of orientations averaged over.

Let us first summarize the transformation of the electron beam that defines the overall map \eqref{eq:map_cm1}. Given a crystal position $\vR_{\perp}$, the function $K(\vr_{\perp},\Omega)$ in \eqref{eq:K1} describes the amplitude of the electron wavefunction at $\vr_{\perp}$ on the detection plane. To arrive there, the incoming wavefunction undergoes three transformation steps. First, it is multiplied by the averaged homogeneous density $\overline{\varrho}$ describing the cross-section area within which the electron enters the crystal and Bragg scattering can take place. This effective aperture smears out the momentum spread of the initial electron wavefunction, which we give in units of wavenumbers by the standard deviation $\Delta k_{\rm in}$. Concretely, the Fourier transform of \eqref{eq:rhoeff_sphere} for an oblate spheroidal crystal volume, $\tilde{\varrho}(\vk_{\perp}) = 3N_{\rm cell} j_1(k_{\perp}R_M)/k_{\perp}R_M$ with $j_1$ a spherical Bessel function, contributes a spread of the order of $\Delta k_{\rm vol} \approx 4.5/R_M$, as per the first zero of $j_1(x)/x$.

Secondly, the smeared wavefunction undergoes Bragg diffraction at the crystal lattice, which splits it into a weighted superposition of momentum-displaced instances, as seen in \eqref{eq:T1}. Since the Bragg momenta are much larger than the incoming wavefunction spread, $g_{hk\ell} \gg \Delta k_{\rm in} , \Delta k_{\rm vol}$, the displaced wavefunctions do not overlap with one another. 

Thirdly, the transmission mask \eqref{eq:M_mask} blocks the undiffracted wavefunction as well as any Bragg-diffracted one that does not overlap with one of the pinholes in momentum space, labeled by $n$. Only those Bragg orders can pass for which $[R(\Omega) \vg_{hk\ell}]_{\perp} \approx (2\pi n/d)\ve_x$

\label{ref:assumptions}
In what follows, we will make four simplifying assumptions in accordance with realistic experimental settings: (i) the initial electron wavefunction is Gaussian, $\la \vr_{\perp}|\psi_{\rm in} \ra \propto e^{-\Delta k_{\rm in}^2 r_{\perp}^2}$, and illuminates a cross-section area greater than the size of the crystal particle, so that $\Delta k_{\rm in} < \Delta k_{\rm vol}$. 
Next, we assume (ii) that the crystal volume is larger than the spread of its centre-of-mass state, so that we can approximate $\overline{\varrho}(\ovr_{\perp}-\vR_{\perp}) \approx \overline{\varrho}(\ovr_{\perp})$ in \eqref{eq:T1}. We introduce the abbreviation $|\overline{\psi}_{\rm in}\ra := \overline{\varrho}(\ovr_{\perp})|\psi_{\rm in}\ra$ for the smeared (and no longer unit-norm) electron wavefunction. Moreover, (iii) the pinhole apertures $M_0$ shall be large compared to the momentum spread of the smeared electron wavefunction (such that diffraction at this aperture is negligible), but small compared to the distance of neighbouring Bragg peaks. Mathematically, $M_0(|\ovp_{\perp}|)|\overline{\psi}_{\rm in}\ra \approx |\overline{\psi}_{\rm in}\ra$ whereas $M_0(|\ovp_{\perp}-\hbar\vg_{hk\ell}|)|\overline{\psi}_{\rm in}\ra \approx 0$ for any $(hk\ell) \neq 0$.
Finally, (iv) we assume that the crystal lattice is brought into a fixed orientation $\Omega_0$ upon scattering. In this orientation, the reciprocal lattice vector of the chosen reference Bragg peak $(h_0k_0\ell_0)$ shall align with the $x$-axis, $R(\Omega_0)\vg_{h_0k_0\ell_0} = (2\pi/d)\ve_x$, such that the $n$-th pinhole of the transmission mask $M$ transmits only the $n$-th diffraction order with relative amplitude $f_n \equiv f_{nh_0,nk_0,n\ell_0}$. 

Using our assumptions, % (i)-(iii), 
the conditional transformation \eqref{eq:map_cm1} of the centre-of-mass state of the crystal particle reduces to 
\begin{align}\label{eq:gratingTrafo_aligned}
   \rho_{\rm cm} &\mapsto \oK(\vr_{\perp},\Omega_0) \rho_{\rm cm} \oK^\da (\vr_{\perp},\Omega_0), \nonumber \\
   \oK(\vr_{\perp},\Omega_0) &\approx i \la \vr_{\perp}|\overline{\psi}_{\rm in}\ra \sum_{n\neq 0} f_n e^{2\pi i n(x-\oX)/d}. 
\end{align}
This transformation describes the transmission through a one-dimensional grating aperture of period $d$ \cite{hornberger2012colloquium}, shifted by the measured electron position $x$---the particle is effectively diffracted by its own crystal lattice. This is enabled by momentum conservation as each Bragg momentum $2\pi\hbar/d$ imparted on the electron necessarily implies the opposite momentum imparted on the particle, inducing Einstein-Podolsky-Rosen-type entanglement between the wavefunctions of the electron and the particle's centre of mass \footnote{Here we assume that the centre-of-mass state is pure, e.g., a ground-state wavepacket released from the trap.}. The above conditional grating transformation \eqref{eq:gratingTrafo_aligned} preserves the purity of the particle wavefunction, but not its norm. In fact, a successful transformation happens with the probability of an electron arriving anywhere on the detection plane,
\begin{align}\label{eq:Pr_det_aligned}
    \Pr{}_{\rm det} &= \int \diff^2 r_{\perp} \, \tr \left\{ \rho_{\rm cm} \oK^\da (\vr_{\perp},\Omega_0) \oK (\vr_{\perp},\Omega_0) \right\} \nonumber \\
    &\approx \la \overline{\psi}_{\rm in} | \overline{\psi}_{\rm in} \ra \sum_{n\neq 0} |f_n|^2 \\
    &= \frac{2N_{\rm cell}^2b_M^2}{V^2} \left[2 -  \frac{1-e^{-2\Delta k_{\rm in}^2 R_M^2}}{\Delta k_{\rm in}^2 R_M^2} \right] \sum_{n\neq 0} |f_n|^2 . \nonumber
\end{align}
where the approximation follows from our assumption (ii) that the Bragg-shifted electron wavefunctions do not overlap.

The theoretical treatment of near-field interference of a free-falling nanoparticle at a periodic grating was already detailed in Ref.~\cite{bateman2014near}; we provide a short re-derivation for the present case in App.~\ref{app:phasespace_calc}. One starts from a (mixed) Gaussian state of the trapped particle with a small position standard deviation $\sigma_X \ll d$ relative to the grating period and a large momentum standard deviation $\sigma_P \gg 2\pi\hbar/d$ relative to the grating momentum, which corresponds to an incoherent point source of matter waves. Upon release, the particle may evolve freely for the time $t_0$ before the effective grating transformation and for the time $t$ after; coherent dispersion over at least one grating period demands that $t_0$ be of the order of the Talbot time, $t_0 \gtrsim T_M = Md^2/2\pi\hbar$. Assumption (ii) remains valid as long as $\sigma_P t_0/M \ll R_M$. Finally, one recaptures the particle and measures its position on the $X$-axis, which one finds to be distributed according to
\begin{align}
    w_3(X) \approx & |\la \vr_{\perp}|\overline{\psi}_{\rm in} \ra |^2\sum_n e^{2\pi i n (X/D-x/d) - 2\pi^2 n^2[\sigma_X t/d(t+t_0)]^2} \nonumber \\
    & \times \sum_j f_j f_{j+n}^* e^{i\pi n (2j+n) tt_0/T_M (t+t_0)} \nonumber \\
    & \times \frac{e^{-[X+(j+n/2)d t/T_M]^2/2\tilde{\sigma}_X^2}}{\sqrt{2\pi} \tilde{\sigma}_X}, \label{eq:qpattern_aligned}%\\
\end{align}
given the measured electron position $\vr_{\perp} = (x,y)$. 
This density, normalised to $\int \diff^2 r_{\perp} \diff X \, w_3(X) \approx \Pr{}_{\rm det}$, exhibits fringes of a geometrically magnified period $D\approx d(t+t_0)/t_0$ and visibility reduced by the finite source size $\sigma_X$, inside a broad Gaussian envelope of width $\tilde{\sigma}_X \approx \sigma_P (t+t_0)/M$. 

Notice that the appearance of a fringe pattern in the near field of a grating does not always certify quantum interference as it could also be explained by a classical shadow effect. Interpreting the initial Gaussian state as the phase-space distribution of a classical particle and treating the grating aperture in \eqref{eq:K1} as a classical shadow mask, we obtain the fringe pattern
\begin{align}
    w_3^{\rm cl}(X) \approx & |\la \vr_{\perp}|\overline{\psi}_{\rm in} \ra |^2 \sum_n e^{2\pi i n (X/D-x/d) - 2\pi^2 n^2[\sigma_X t/d(t+t_0)]^2} \nonumber \\
    & \times \sum_j f_j f_{j+n}^* \frac{e^{-X^2/2\tilde{\sigma}_X^2}}{\sqrt{2\pi} \tilde{\sigma}_X}. \label{eq:clpattern_aligned}
\end{align}
The appearance of fringes according to the quantum prediction \eqref{eq:qpattern_aligned} does not signify matter-wave coherence whenever they match the classical prediction \eqref{eq:clpattern_aligned}. This is for instance the case when $t \ll T_M$. In the regime of a very broad Gaussian envelope, $\tilde{\sigma}_X \gg dt/T_M$, it turns out that the quantum and classical patterns always agree whenever the grating aperture in \eqref{eq:K1} comprises only one pair of opposite Bragg orders, say, $n=\pm 1$. For a clear quantum signature, one must therefore select at least two Bragg orders of different magnitude; see App.~\ref{app:phasespace_calc}.

\section{Phase-space description of the matter-wave diffraction scheme}\label{app:phasespace_calc}

Our setting is a nanoparticle initially in a Gaussian state, which is then released and diffracted at an effective grating, generated by electron diffraction at the nanoparticle with subsequent filtering through an array of pinholes in the Fourier plane and position-resolved detection in the image plane. Conditioned on the detected electron position $(x,y)$, the effective grating transformation of the center-of-mass state is given by
\begin{align}
    \rho_{\rm cm} \mapsto \sum_{n,n'} B_{n,n'}^{(x,y)} e^{2\pi i n(x-\oX)/d}\rho_{\rm cm}e^{2\pi i n'(\oX-x)/d}, \label{eq:gratingTrafo_rho_both}
\end{align}
where the coefficients $B_{n,n'}^{(x,y)}$ depend on the orientation state of the crystal particle, as derived in App.~\ref{app:gratingTrafo}. Since only the motional state of the particle along the $\ve_x$-axis of the pinhole array matters, we can adopt the one-dimensional near-field interference formalism of Ref.~\cite{bateman2014near} and replace the periodic phase grating transformation there by the above transformation. 

In the phase-space formalism, we represent the one-dimensional center-of-mass state in terms of the Wigner function,
\begin{equation}
    w(X,P) = \int \frac{\diff s}{2\pi \hbar} e^{iPs/\hbar} \left\la X-\frac{s}{2} \right| \rho_{\rm cm} \left| X + \frac{s}{2} \right\ra .
\end{equation}
We will assume that the initial state of the nanoparticle upon release is a Gaussian of arbitrary widths $\sigma_X,\sigma_P$ obeying $\sigma_X\sigma_P \geq \hbar/2$. The respective Wigner function is
\begin{equation}
    w_0(X,P) = \frac{1}{2\pi \sigma_X \sigma_P} e^{-X^2/2\sigma_X^2 - P^2/2\sigma_P^2}. \label{eq:w0}
\end{equation}
In the case of a harmonic oscillator ground state, we have $\sigma_X = \sqrt{\hbar/2M\omega}$ and $\sigma_P = \hbar/2\sigma_X = \sqrt{\hbar M\omega/2}$, given the trapping frequency $\omega$. For a thermal state of finite temperature, one multiplies both $\sigma_X$ and $\sigma_P$ by $\sqrt{\coth(\hbar\omega/2k_B T)}$. For the temperature considered in the main text, thermal corrections to the ground state widths are quite small. 

The conditional grating transformation \eqref{eq:gratingTrafo_rho_both} can be represented in phase space in terms of a momentum convolution of the Wigner function, $w(X,P) \mapsto \int \diff Q\, \cT^{(x,y)} (X,P-Q) w(X,Q)$, with the kernel
\begin{align}
    \cT^{(x,y)}(X,P) &= \sum_{j,n} e^{2\pi i n (X-x)/d} B_{j,j+n}^{(x,y)} \delta \left( P + \frac{(2j+n)\pi \hbar}{d} \right) \nonumber \\
    &= \sum_{n} e^{2\pi i n (X-x)/d} \int \frac{\diff s}{2\pi \hbar} e^{iPs/\hbar} B_n^{(x,y)} \left( \frac{s}{d}\right), \label{eq:condGrating_kernel}
\end{align}
where we introduced the so-called Talbot coefficients as in \cite{bateman2014near},
\begin{equation}
    B_n^{(x,y)} (\xi) = \sum_j B_{j,j+n}^{(x,y)} e^{i\pi\xi(n+2j)} = B_{-n}^{(x,y)*}(-\xi), \label{eq:TalbotCoeff} 
\end{equation}
The latter identity holds for $\xi \in \mathbb{R}$ and guarantees that the Wigner function is real-valued; it follows straightforwardly from the fact that the transformation coefficients obey $B_{j,k}^{(x,y)} = B_{k,j}^{(x,y)*}$.

One of the key questions for us is whether the observed fringe pattern is non-classical. 
A direct and setup-specific way is to compare the interferogram predicted by \eqref{eq:gratingTrafo_rho_both} to the shadow fringe pattern obtained by treating the effective grating as a classical aperture in phase space. This is achieved by omitting the argument of the Talbot coefficients \eqref{eq:TalbotCoeff} and setting it to zero, so that the transformation kernel \eqref{eq:condGrating_kernel} no longer imprints diffraction momenta onto the Wigner function and the transformation reduces to the density modulation caused by an effective classical transmission mask with Fourier components $B_n^{(x,y)}(0)$,
\begin{align}
    \cT_{\rm cl}^{(x,y)}(X,P) &=  \delta(P) \sum_{n} e^{2\pi i n (X-x)/d} B_n^{(x,y)} (0) \label{eq:condGrating_kernel_cl}.
\end{align}
Such a classical mask will also produce periodic fringes in the near field. Conversely, if the quantum interferogram cannot be distinguished from this classical fringe pattern (whether the quantum calculation results in Wigner negativities or not), then one cannot rule out a simple classical model and claim to observe a genuine quantum diffraction effect. We employ this criterion for the interferograms we evaluate here.

For the Wigner function right after the grating transformation, we first propagate the initial state \eqref{eq:w0} freely by the time $t_0$ and then apply the grating transformation \eqref{eq:condGrating_kernel},
\begin{align}
    w_2(X,P) =& \int \diff Q \, \cT^{(x,y)}(X,P-Q) w_0 \left( X - \frac{Qt_0}{M},Q \right) \nonumber \\
    =& \sum_{j,n} \frac{B_{j,j+n}^{(x,y)}}{2\pi \sigma_X \sigma_P}  e^{- [X-Pt_0/M-(2j+n)dt_0/T_M]^2/2\sigma_X^2} \nonumber \\
    &\times e^{2\pi i n (X-x)/d -[P+(2j+n)h/2d]^2/2\sigma_P^2},
\end{align}
where $T_M = Md^2/2\pi\hbar$ denotes the Talbot time. 

For the final fringe pattern, we propagate the Wigner function $w_2$ freely for another time $t$ and then obtain the position distribution by integrating over the momentum coordinate. We arrive at
\begin{align}
    w_3(X) &= \int \diff P \, w_2 \left( 
X-\frac{Pt}{M},P \right) \nonumber \\
&= \sum_{j,n} B_{j,j+n}^{(x,y)} e^{2\pi i n (X-x)/d}  A_{n,j+n/2} (X). 
\end{align}
Here, we introduce the abbreviation $A_{n,k} (X)$ for the Fourier integral of the Gaussian initial Wigner function over $P$. 
The classical fringe pattern resulting from the kernel \eqref{eq:condGrating_kernel_cl} is obtained by replacing $A_{n,k}(X) \to A_{n,0} (X)$.
Explicitly, we have
\begin{widetext}
\begin{align}
    A_{n,k} (X) &= \int \frac{\diff P}{2\pi \sigma_X \sigma_P} e^{-ind t P/\hbar T_M - (P+kh/d)^2/2\sigma_P^2} \, e^{-[X-P(t+t_0)/M - kd t_0/T_M]^2/2\sigma_X^2} \nonumber \\
    &= e^{2\pi i n k t/T_M}  \int \frac{\diff P}{2\pi \sigma_X \sigma_P} e^{-ind t P/\hbar T_M - P^2/2\sigma_P^2} \, e^{-[X-P(t+t_0)/M+ kdt/T_M]^2/2\sigma_X^2}
\end{align}
To simplify the Gaussian terms in the exponent, which we need to integrate over $P$, we define the auxiliary rescaled position and momentum widths,
\begin{align}
    \tilde{\sigma}^2_P &= \frac{1}{\frac{1}{\sigma_P^2} + \frac{(t+t_0)^2}{M^2\sigma_X^2} } = \frac{\sigma_P^2}{1+\left[ \frac{\sigma_P(t+t_0)}{M\sigma_X} \right]^2}, \label{eq:sigP_prop}\\
    \tilde{\sigma}^2_X &= \frac{\sigma_X^2}{1-\frac{\tilde{\sigma}_P^2 (t+t_0)^2}{M^2\sigma_X^2}} = \sigma_X^2 + \left[\frac{\sigma_P (t+t_0)}{M}\right]^2 , \label{eq:sigX_prop}
\end{align}
which preserve $\tilde{\sigma}_X\tilde{\sigma}_P = \sigma_X\sigma_P$, as one can easily check. Abbreviating temporarily $X_k = X + kdt/T_M$, the Gaussian terms in the exponent can then be expanded as
\begin{align}
    -\frac{P^2}{2\sigma_P^2} - \frac{[X_k - P(t+t_0)/M]^2}{2\sigma_X^2}
    &= - \frac{P^2}{2\tilde{\sigma}_P^2} - \frac{X_k^2}{2\sigma_X^2} + 2 \frac{PX_k(t+t_0)}{2 M \sigma_X^2} \nonumber \\
    &= - \frac{X_k^2}{2\sigma_X^2} - \frac{[P-\tilde{\sigma}_P^2 (t+t_0)X_k/M\sigma_X^2]^2}{2\tilde{\sigma}_P^2} + \frac{\tilde{\sigma}_P^2 (t+t_0)^2 X_k^2}{\sigma_X^4 M^2} \nonumber \\
    &= - \frac{X_k^2}{2\tilde{\sigma}_X^2} -  \frac{[P-\tilde{\sigma}_P^2 (t+t_0)X_k/M\sigma_X^2]^2}{2\tilde{\sigma}_P^2}.
\end{align}
This shifted Gaussian can be straightforwardly integrated over $P$ now, resulting in
\begin{align}
    A_{n,k} (X) &= \frac{e^{-X_k^2/2\tilde{\sigma}_X^2}}{\sqrt{2\pi} \tilde{\sigma}_X} e^{-\tilde{\sigma}_P^2 (n d t/\hbar T_M)^2/2} \, e^{2\pi i n (t/T_M)[k - \tilde{\sigma}_P^2 X_k d (t+t_0)/M h \sigma_X^2]} \nonumber \\
    &=  \frac{e^{-X_k^2/2\tilde{\sigma}_X^2}}{\sqrt{2\pi} \tilde{\sigma}_X} R_n e^{2\pi i n k t d/T_M D + 2\pi i n X (d/D-1)/d}.
\end{align}
Here, we introduced the $n$-th order fringe reduction factor $R_n$ and what will become the magnified fringe period $D$, 
\begin{align}
    R_{n} &= R_{-n} = e^{-\tilde{\sigma}_P^2 (n d t/\hbar T_M)^2/2} = e^{-2\pi^2 n^2 (\tilde{\sigma}_P t/Md)^2} , \\
    D &= \frac{d}{1 - \frac{\tilde{\sigma}_P^2 t(t+t_0)}{M^2\sigma_X^2}} = d \frac{(t+t_0)^2 + (M\sigma_X/\sigma_P)^2}{t_0(t+t_0) + (M\sigma_X/\sigma_P)^2} .
\end{align}
Now we can put everything together and obtain the quantum interferogram and the classical fringe pattern,
\begin{align}
    w_3(X) &= \sum_n e^{2\pi i n X/D - 2\pi i nx/d} R_n \sum_j B_{j,j+n}^{(x,y)} e^{i\pi n (2j+n)td/T_M D} \frac{e^{-[X+(j+n/2)d t/T_M]^2/2\tilde{\sigma}_X^2}}{\sqrt{2\pi} \tilde{\sigma}_X}, \label{eq:w3_q_full} \\
    w_3^{\rm cl} (X) &= \frac{e^{-X^2/2\tilde{\sigma}_X^2}}{\sqrt{2\pi} \tilde{\sigma}_X} \sum_n e^{2\pi i n X/D - 2\pi inx/d} R_n B_n (0). \label{eq:w3_cl_full} 
\end{align}
These are the expressions we use to evaluate the fringe patterns in the main text. For an appreciable fringe visibility, we need to have sizeable $R_{n\neq 0}$, implying $[\sigma_X t/d(t+t_0)] \ll 1$. At the same time, we also aim for a strong fringe magnification, implying that $t \gg t_0$ so that $D \approx dt/t_0 \gg d$. The consequence is that we need to have $\sigma_X \ll d$ for visible fringes. A tradeoff must be chosen carefully here. 

In order to highlight the regime in which the quantum and the classical fringe patterns coincide approximately, we consider the scenario in which the nanoparticle is initially well localized, $\sigma_X \lesssim d$, but has a broad momentum distribution compared to the diffraction scale, $\sigma_P \gg 2\pi\hbar/d$. This implies that $\sigma_P (t+t_0)/M\sigma_X \gg (d/\sigma_X)(t+t_0)/T_M \gtrsim 1$ for the relevant $t,t_0$, and we may then approximate 
\begin{equation}
    \tilde{\sigma}_X \approx \frac{\sigma_P (t+t_0)}{M}, \quad \tilde{\sigma}_P \approx \frac{M \sigma_X}{t+t_0} \qquad \Rightarrow \qquad D \approx d \frac{t+t_0}{t_0}, \quad R_n \approx e^{-2\pi^2 n^2 \sigma_X^2t^2/d^2(t+t_0)^2}. 
\end{equation}
These approximations are already applied in the above expressions \eqref{eq:qpattern_aligned} and \eqref{eq:clpattern_aligned}.
If we can also neglect the small position displacements (of order $d$) in the broad Gaussian envelope of \eqref{eq:w3_q_full}, we obtain the quantum expression
\begin{align}
     w_3 (X) &\approx \frac{e^{-[MX/\sigma_P (t+t_0)]^2/2}}{\sqrt{2\pi} \sigma_P (t+t_0)/M}  \sum_n e^{2\pi i n X/D + in\phi_x} R_n B_n \left[ n \frac{tt_0}{T_M(t+t_0)} \right]. \label{eq:w3_q_broad}  
\end{align}
For the corresponding classical version, simply set the argument of $B_n[\ldots] \to B_n (0)$.
Comparing the final expression \eqref{eq:w3_q_broad} to its classical counterpart tells us that we need to have $B_{n\neq 0} (\xi) \neq \text{const}$ for non-classical behaviour in this approximated scenario. \emph{This implies that we need to condition on more than a single pair of $\pm N$-th diffraction orders for non-classical fringe terms to occur.}

We prove this statement by  assuming that we condition on only the $\pm N$-th diffraction order, i.e., the only non-zero grating coefficients are $B_{-N,-N}^{(x,y)},B_{N,N}^{(x,y)},B_{-N,N}^{(x,y)},B_{N,-N}^{(x,y)}$. Then, 
\begin{align}
    B_n (n\xi) &= \delta_{n0} \left[ B_{-N,-N}^{(x,y)} + B_{N,N}^{(x,y)} \right] + \delta_{n,2N} B_{-N,N}^{(x,y)} e^{i\pi n\xi (n-2N)} + \delta_{n,-2N} B_{N,-N}^{(x,y)} e^{i\pi n\xi (n+2N)} \nonumber \\
    &= \delta_{n0} \left[ B_{-N,-N}^{(x,y)} + B_{N,N}^{(x,y)} \right] + \delta_{n,2N} B_{-N,N}^{(x,y)} + \delta_{n,-2N} B_{N,-N}^{(x,y)} = B_n (0).
\end{align}
Hence the quantum and the classical fringe patterns are indistinguishable. 

Quantum behaviour arises if we allow for mixed terms between at least two different diffraction orders, e.g., we condition on $\pm 1$ and $\pm 2$. This implies that genuine quantum interference is more prominent the greater odd-integer Talbot coefficients such as $B_{\pm 1}, B_{\pm 3}$ are in magnitude. 

\section{Estimation of macroscopicity}
\label{app:macroscopicity}

The empirical macroscopicity of a mechanical quantum  experiment is based on the extent to which the observation of nonclassical phenomena rules out a class of macrorealistic modifications of quantum theory \cite{nimmrichter2013macroscopicity}. In centre-of-mass interference experiments, this class of modifications causes a loss of quantum coherence similar to conventional decoherence processes, which must be included in the free time evolution of the particle before and after the effective grating interaction. 

Explicitly, the modification contributes a Lindblad dissipator to the free evolution of the reduced centre-of-mass state, $ \partial_t \rho_{\rm cm} = -i[\oP^2/2M\hbar,\rho_{\rm cm}] + \cL \rho_{\rm cm} $, the position representation of which reads as
\begin{equation}\label{eq:macro_Gamma}
    \la X| \cL\rho_{\rm cm}|X'\ra = - [\Gamma (0) - \Gamma(X-X')]  \la X| \rho_{\rm cm}|X'\ra, \qquad \Gamma(X) = \frac{1}{m_0^2 \tau_0} \int \frac{\diff^3 q}{(2\pi\sigma_q^2)^{3/2}} e^{-q^2/2\sigma_q^2}| \tilde{\varrho}_M (\vq)|^2 e^{-iq_x X}.
\end{equation}
Here, $\tau_0$ is a time parameter, $\sigma_q$ a momentum width parameter (here in units of wavenumbers), and $m_0$ is a reference mass taken to be that of the electron in previous comparative studies. Accordingly, we also restrict to super-atomic momentum widths, $1/\sigma_q \gtrsim 1\,$nm. The integrand contains the Fourier transform of the particle's mass density. Assuming a homogeneous mass density for our disc-shaped spheroidal particle of volume $V=4\pi R_M^2 b_M/3$, the Fourier transform can be shown to take the simple form
\begin{equation}
    \tilde{\varrho}_M (\vq) = \frac{M}{V} \int_V \diff^3 r\, e^{-i\vq\cdot\vr}= 3M \frac{j_1 \left[ \sqrt{ (q_{\perp} R_M)^2 + (q_z b_M)^2} \right]}{\sqrt{ (q_{\perp} R_M)^2 + (q_z b_M)^2}},
\end{equation}
where $j_1$ is a spherical Bessel function and $q_{\perp} = \sqrt{q_x^2+q_y^2}$. Noting that the function vanishes for $q_{\perp} \gg 1/R_M$ and that the relevant spatial coherences are confined to the atomic scale in our case, $|X-X'| \lesssim d$, we can Taylor-expand the Fourier exponential in \eqref{eq:macro_Gamma}. To lowest non-vanishing order, we obtain
\begin{align} \label{eq:Gamma_approx}
\Gamma(0) - \Gamma(X-X') 
    &\approx \frac{9\pi M^2 (X-X')^2}{2m_0^2 \tau_0} \int_{-\infty}^\infty \diff q_z \int_0^{\infty}\diff q_{\perp}\, q_{\perp}^3 \frac{e^{-(q_{\perp}^2+q_z^2)/2\sigma_q^2}}{(2\pi\sigma_q^2)^{3/2}} \frac{j_1 \left[ \sqrt{ (q_{\perp} R_M)^2 + (q_z b_M)^2} \right]}{\sqrt{ (q_{\perp} R_M)^2 + (q_z b_M)^2}} \\
    &= \frac{9M^2 \sigma_q^2 (X-X')^2}{2\sqrt{2\pi} m_0^2 \tau_0} \mathcal{I} (\sigma_q R_M,\sigma_q b_M), \quad \mathcal{I} (\alpha,\beta) = \iint_0^\infty \diff \xi_{\perp} \diff \xi_z \, \xi_{\perp}^3 e^{-(\xi_{\perp}^2+\xi_z^2)/2} \frac{j_1\left(\sqrt{\alpha^2\xi_{\perp}^2 + \beta^2\xi_z^2}\right)}{\sqrt{\alpha^2\xi_{\perp}^2 + \beta^2\xi_z^2}} .\nonumber 
\end{align}
The remaining double integral $\mathcal{I}$ can be evaluated numerically for each argument $(\alpha,\beta)$. 

The contribution of the dissipator \eqref{eq:macro_Gamma} to the free evolution is most conveniently expressed in the characteristic function representation of the Wigner function. This was done in Ref.~\cite{bateman2014near} in the limit of an arbitrarily incoherent initial particle state, $\sigma_P \to \infty$, which remains valid here as long as the net broadening of the Gaussian envelope due to \eqref{eq:macro_Gamma} is negligible. This is indeed the case here, since we assume $\sigma_P \gg 2\pi\hbar/d > \hbar \sigma_q$. As a result, the reduction factors $R_n$ in the quantum fringe pattern \eqref{eq:w3_q_full} are simply multiplied by another exponential decay term,
\begin{equation}
    R_n \to R_n \exp \left\{ (t+t_0) \int_0^1 \diff \vartheta \left[ \Gamma \left( \frac{n d t t_0 \vartheta}{T_M (t+t_0)} \right) - \Gamma(0)\right] \right\} \approx R_n \exp \left\{-\frac{3M^2 (t+t_0)}{2\sqrt{2\pi} m_0^2 \tau_0} \left[\frac{n\sigma_q d t t_0}{T_M (t+t_0)}\right]^2 \mathcal{I} (\sigma_q R_M,\sigma_q b_M) \right\}.
\end{equation}
The macroscopicity $\mu$ of a quantum experiment is then given by the greatest value of the time parameter $\tau_0$, maximized with respect to $\sigma_q$, that is ruled out by an observation of quantum interference fringes. For our proposed case study, we assume that one could observe more than $50\%$ of the predicted quantum fringe contrast of order $n=2$ at $t=1\,$ms and $t_0=T_M$, as depicted in the bottom end of Fig.~2(a) in the main text. Given $\sigma_q$, this rules out
\begin{equation}
    \tau_0 \leq \tau_{\max} (\sigma_q) = \frac{6(M/m_0)^2 }{\sqrt{2\pi} \ln 2 } \frac{t^2}{t+T_M} (\sigma_q d)^2 \mathcal{I} (\sigma_q R_M,\sigma_q b_M).
\end{equation}
We numerically maximize to arrive at $\mu = \log_{10} \max_{\sigma_q} \tau_{\max} (\sigma_q) \approx 16.3$.

\section{Effect of imperfect orientational alignment}\label{app:rotationAverage}

Here we estimate the loss of coherence between the Bragg orders due to a finite uncertainty in the orientation of the nanocrystal. To this end, we are going to assume that the nanocrystal is sufficiently aligned so that only the selected family of Bragg orders passes through the pinholes and all others are blocked, $(hk\ell) = n(h_0k_0\ell_0) $. As before, we shall also assume that the spread of the center-of-mass state is small compared to the nanocrystal size.  
Let $\Omega_1$ be the deviation of the orientation from the ideal alignment $\Omega_0$, $R(\Omega) = R(\Omega_1)R(\Omega_0)$ with $R(\Omega_0)\vg_{hk\ell} = (2\pi n/d)\ve_x$, and let $\mu(\Omega_1)$ be a narrow distribution of such deviations. The main effect is then a modulation of the electron wavefunction passing the transmission mask. We can replace the ideal grating transformation \eqref{eq:gratingTrafo_aligned} by 
\begin{align}
    \rho_{\rm cm} &\mapsto \int\diff \Omega_1 \, \mu(\Omega_1) \oK(\vr_{\perp},\Omega_1) \rho_{\rm cm} \oK^\da(\vr_{\perp},\Omega_1) , \\  
    \oK(\vr_{\perp},\Omega_1) &\approx i \la \vr_{\perp}| \sum_{n\neq 0} f_n M_0 \left(\left|\ovp_{\perp} - \frac{nh}{d} \ve_x\right|\right) e^{2\pi i n(\ox -\oX)/d} e^{2\pi i n\ovr_{\perp} \cdot [R(\Omega_1)-\id]\ve_x/d} | \overline{\psi}_{\rm in}\ra \nonumber \\
    &= i \sum_{n\neq 0} f_n  e^{2\pi i n(x-\oX)/d} e^{2\pi i n\vr_{\perp} \cdot [R(\Omega_1)-\id]\ve_x/d} \la \vr_{\perp}| M_0 \left( \left| \ovp_{\perp} + \frac{nh}{d} [R(\Omega_1)-\id]\ve_x \right|\right) | \overline{\psi}_{\rm in}\ra \nonumber \\
    &\approx i \sum_{n\neq 0} f_n  e^{2\pi i n(x-\oX)/d} e^{2\pi i n\vr_{\perp} \cdot [R(\Omega_1)-\id]\ve_x/d} M_0 \left( \frac{nh}{d}\left| [R(\Omega_1)-\id]\ve_x \right|\right) \la \vr_{\perp} | \overline{\psi}_{\rm in}\ra . \nonumber
\end{align}
In the third line, we make use of the canonical commutation relations to commute the mask aperture function $M_0$ with the momentum displacement operators. In the fourth line, we then make use of assumption (iii) that the smeared wavefunction $|\overline{\psi}_{\rm in}\ra$ is narrow in momentum space compared to the pinhole aperture, while allowing for a larger alignment spread (see page \pageref{ref:assumptions}). 

The now only partially coherent grating transformation  has the form \eqref{eq:gratingTrafo_rho_both}, with coefficients $B_{n,n'}^{(x,y)} = |\la \vr_{\perp}|\overline{\psi}_{\rm in}\ra|^2 f_n f_{n'}^* D_{n,n'}(\vr_{\perp})$ and
\begin{equation}
    D_{n,n'}(\vr_{\perp}) \approx \int\diff^3\Omega_1 \,\mu(\Omega_1) e^{2\pi i (n-n')\vr_{\perp} \cdot [R(\Omega_1)-\id]\ve_x/d} M_0 \left( \frac{nh}{d}\left| [R(\Omega_1)-\id]\ve_x \right|\right) M_0^* \left( \frac{n'h}{d}\left| [R(\Omega_1)-\id]\ve_x \right|\right). 
\end{equation}
These terms describe the loss of coherence between two different Bragg orders $n\neq n'$ due to orientational averaging. They also describe an additional loss of signal if the orientational spread exceeds the pinhole in size, so that $D_{n,n} < 1$. 
For the sake of simplicity, let us assume Gaussian pinhole apertures of relative width $\xi$, $M_0 (z h/d) = \exp(-z^2/2\xi^2)$.

In the regime of small deviations from perfect alignment, we can use the Euler angle representation in the XZX convention and approximate to second order,
\begin{equation}
    [R(\Omega_1)-\id]\ve_x = [R_x(\gamma) R_z(\beta)-\id ]\ve_x 
    %= R_x(\gamma) [\cos\beta \ve_x + \sin\beta \ve_y] - \ve_x 
    = [\cos\beta -1]\ve_x + \sin\beta \cos\gamma \ve_y + \sin\beta \sin\gamma \ve_z \approx -\frac{\beta^2}{2} \ve_x + \beta \ve_y + \beta\gamma \ve_z.
\end{equation}
To lowest (first) order, only the nutation angle $\beta$ is  significant and we get
\begin{equation}
    D_{n,n'} (\vr_{\perp}) \approx \int_{-\pi/2}^{\pi/2} \diff \beta \, \overline{\mu} (\beta) e^{-(n^2 +n'^2)\beta^2/2\xi^2} e^{2\pi i (n-n') y \beta/d} \approx \frac{\xi}{\sqrt{\xi^2+(n^2+n'^2)\sigma_\beta^2}} \exp \left[ - \frac{2\pi^2(n-n')^2\sigma_\beta^2\xi^2}{\xi^2 + (n^2+n'^2)\sigma_\beta^2} \, \frac{y^2}{d^2}\right],
\end{equation}
given an approximately Gaussian spread of nutation angles with standard deviation $\sigma_{\beta} \ll \pi/2$. Recall that the electron position is approximately confined to the crystal volume, $y \lesssim R_M$. 
In the case of large pinholes, $\xi \gg \sigma_\beta$,
coherence loss due to orientational averaging thus becomes detrimental when $\sigma_{\beta} \gtrsim d/\pi R_M \sim 10^{-3}$---posing a very stringent alignment requirement. 
One alleviates this problem by using narrow pinholes, $\xi \ll \sigma_\beta$, so that
\begin{equation}\label{eq:Dnn_smallPinholes}
    D_{n,n'} (\vr_{\perp}) \approx \frac{\xi}{\sigma_\beta \sqrt{n^2+n'^2}} \exp \left[ -2 \frac{\pi^2 \xi^2 y^2}{d^2} \frac{(n-n')^2}{n^2+n'^2} \right] .
\end{equation}
Interference remains largely visible for $\xi \sim d/\pi R_M$. Smaller $\xi$-values would be even better, but they violate our requirement (iii) that the pinholes be large compared to the momentum spread of the diffracted electron wavefunction.

\begin{figure}
    \centering
    \includegraphics[width=1.0\linewidth]{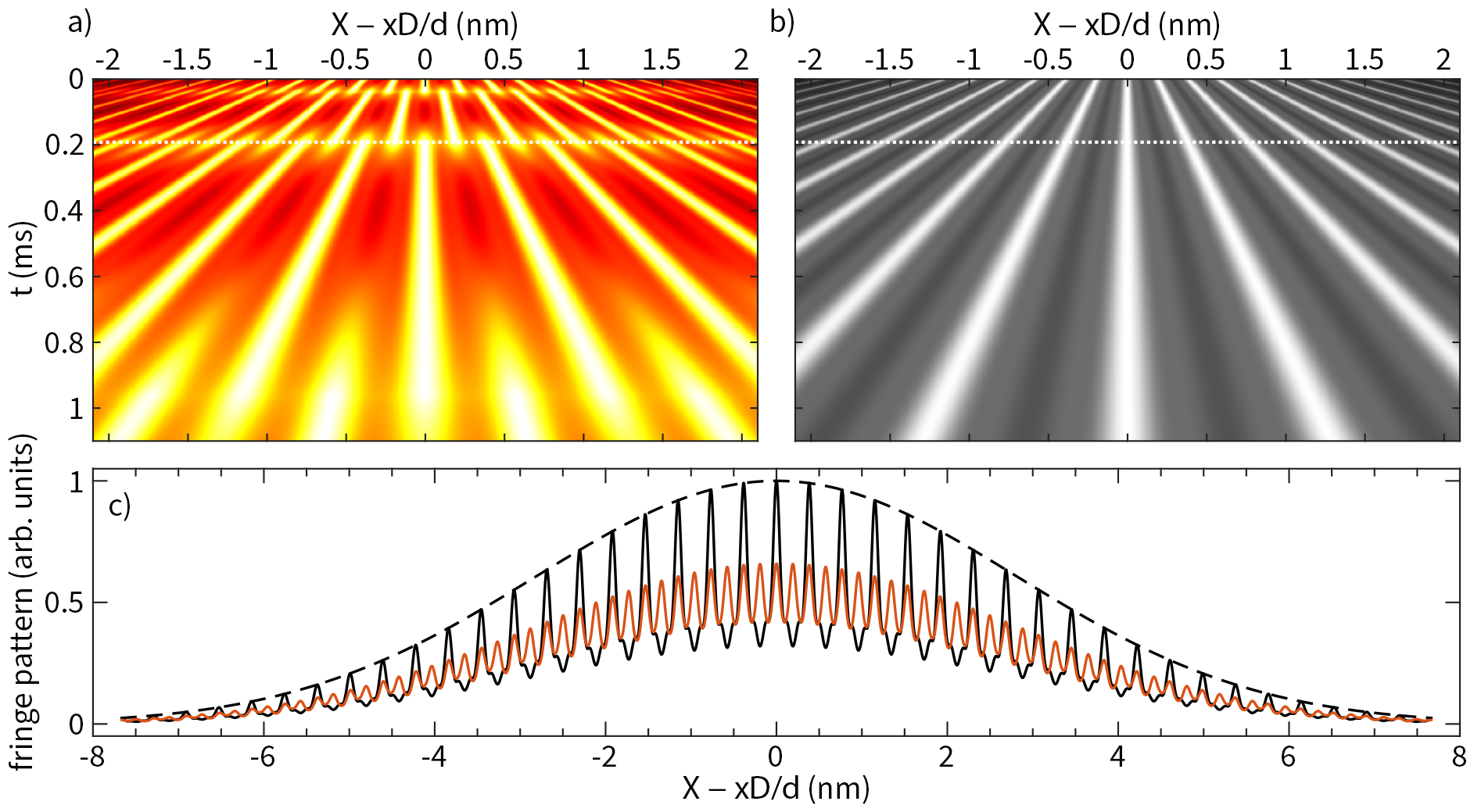}
    \caption{Fringe patterns for an imperfectly aligned silicon nanocrystal of mass $M=2\times 10^9\,$amu, based on the  settings of Fig.~2 in the main text. Rather than perfect alignment, we here assume that the crystal orientation is uncertain, but filtered through narrow Gaussian pinhole apertures ($\xi=10^{-3}$) of the Bragg transmission mask. The fringe patterns shown here are conditioned on the detection of the electron at $y=R_M/2$. Detection at smaller (greater) $|y|$ results in higher (lower) fringe contrast. It is most likely to find $|y| \lesssim R_M$.}
    \label{fig:carpet_decoh}
\end{figure}

For a ballpark estimate based on the exemplary setup in the main text, let $\xi=10^{-3} \ll \sigma_\beta$ such that \eqref{eq:Dnn_smallPinholes} applies and assume, quite conservatively, that the electron was detected at $y=R_M/2$. Keeping all other parameters the same as in Fig.~2 of the main text, the expected fringe patterns for the nanoparticle are shown in Fig.~\ref{fig:carpet_decoh}.

\end{widetext}

\end{document}